\documentclass{aa}
\usepackage{epsfig}
\begin{document}

\title{Str\"omgren $uvby$ photometry of the open clusters NGC~6192 and NGC~6451
\thanks{Based on observations at the Cerro Tololo Inter-American Observatory
which is operated by the Association of Universities for Research in Astronomy 
Inc. (AURA), under a cooperative agreement with the National Science Foundation 
as part of the National Optical Astronomy Observatories.}}

\author{E.~Paunzen\inst{1,2}, H.M.~Maitzen\inst{1}, K.D.~Rakos\inst{1},
J.~Schombert\inst{3}}

\mail{Ernst.Paunzen@univie.ac.at}

\institute{Institut f\"ur Astronomie der Universit\"at Wien,
           T\"urkenschanzstr. 17, A-1180 Wien, Austria
\and       Zentraler Informatikdienst der Universit\"at Wien,
           Universit\"atsstr. 7, A-1010 Wien, Austria
\and	   Department of Physics, University of Oregon, Eugene,
		   OR 97403, United States of America}

\date{Received 2002; accepted 2002}
\titlerunning{Str\"omgren photometry of NGC~6192 and NGC~6451}{}

\abstract{We have investigated the two open clusters NGC~6192 and
NGC~6451 for which widely different reddening values and thus
ages and distances can be found in the literature via Str\"omgren
$uvby$ photometry. Our measurements allow to disentangle the
apparent discrepancies from the literature and to derive new accurate
values. From appropriate calibrations we find that the overall abundance
for NGC~6192 is about solar whereas a subsolar value for NGC~6451 
was estimated. From two previous reported photometrically candidate
CP stars, one within NGC~6192 shows Str\"omgren indices typical for
a B8\,Si star whereas the other object of NGC~6451 is most probably
a foreground G-type star.
}

\maketitle

\section{Introduction}

The two open clusters NGC~6192 and NGC~6451 are good examples of how
different estimates of the interstellar reddening influence the derived
distances and ages. For NGC~6451, Svolopoulos (1966)  
lists a reddening $E(B-V)$\,=\,0.08\,mag whereas Kjeldsen \& Frandsen (1991)
give $E(B-V)$\,=\,0.70\,mag as best value resulting in a difference of 560\,pc
versus 2100\,pc and 5\,Gyr versus 0.2\,Gyr, respectively. The situation for
NGC~6192 is almost identical. 

We have observed these two open clusters within the Str\"omgren $uvby$ system
in order to derive the reddening as well as the overall metallicity. We have
not tried to independently estimate the ages since no reliable isochrones
within the Str\"omgren $uvby$ system have been published so far. Since no
$\beta$ measurements were made, the calibration of effective temperatures
is not possible on solid grounds. However, ages have been derived in the literature
for both clusters depending on the individual reddening values 
(Kjeldsen \& Frandsen 1991).

Paunzen \& Maitzen (2002) found one photometric CP candidate in each program
cluster via $\Delta a$ photometry (Maitzen 1976). We have investigated
the Str\"omgren indices for these two objects in order to clarify their
nature.

\section{Observations and reduction}

Observations of the two open clusters and several standard
stars were performed 
with the 150\,cm telescope at the Cerro Tololo Inter-American
Observatory (observer: H.M.~Maitzen). With a focus of f/7.5 
and a SITe\,2k CCD (0.44$\arcsec$/pixel), a field of view of 
about 15$\arcmin$ was achieved.

We have used a standard Str\"omgren $uvby$ filter set.
The observing log is listed in Table \ref{log}.
In total, 26 frames for the two clusters in four filters were observed
and used for the further analysis.

The basic reductions (bias-subtraction, dark-correction, 
flat-fielding) were carried out within standard IRAF routines.
For all frames we have applied a point-spread-function-fitting within the
IRAF task DAOPHOT (Stetson 1987). 
Photometry of each frame was performed separately and the measurements 
(corrected for the extinction) were then 
averaged and weighted by their individual photometric error.

\begin{figure*}
\begin{center}
\centerline{\vbox{\psfig{figure=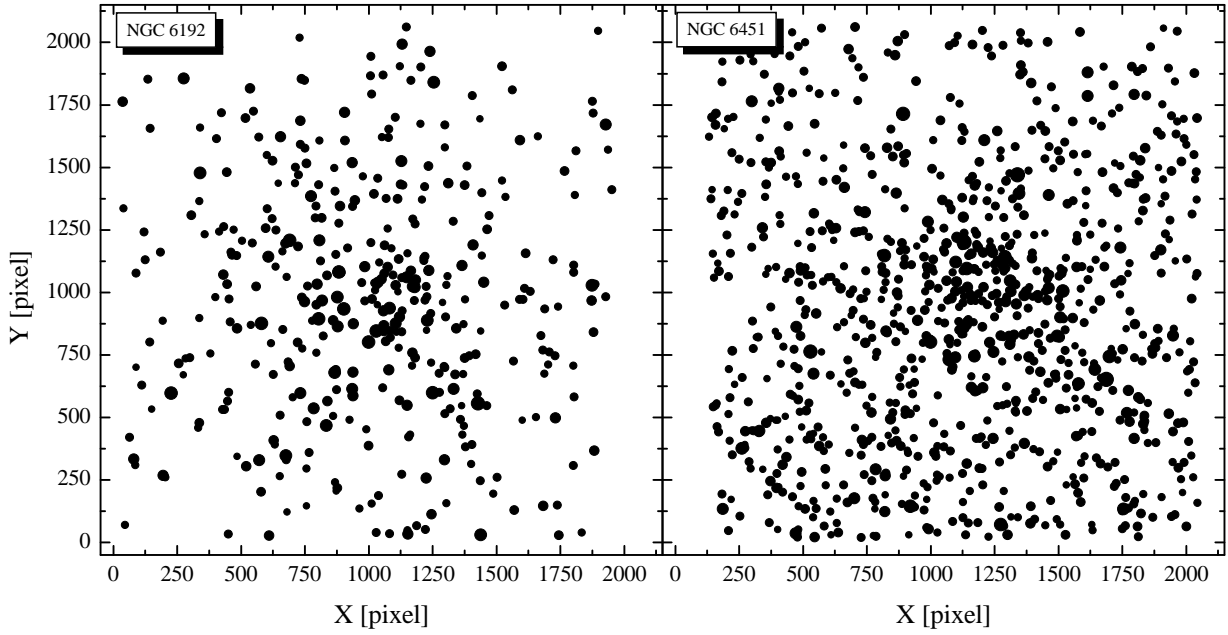,width=170mm}}}
\caption[]{
Finding charts for the program clusters.
North is to the left and west
is upwards; 1 pixel\,=\,0.43$\arcsec$. The sizes (by area) of the open circles
are inversely proportional to the $V$-magnitudes taken from Tables 2 and 3
in the sense that larger open circles denote brighter objects.}
\label{charts}
\end{center}
\end{figure*}

\begin{table}[t]
\begin{center}
\caption{Observing log}
\label{log}
\begin{tabular}{lccccc}
\hline
\hline
Cluster & Night & \#$_{u}$ & \#$_{v}$ & \#$_{b}$ & \#$_{y}$ \\
\hline
NGC~6192 & 02/03.09.2002 & 2 & 2 & 2 & 2 \\
NGC~6451 & 02/03.09.2002 & 2 & 2 & 2 & 2 \\
         & 03/04.09.2002 & 2 & 3 & 3 & 2 \\
\hline
\end{tabular}
\end{center}
\end{table}

The finding charts of our open clusters are shown in Fig. \ref{charts}.
The size of the symbols (by area) is inversely proportional to the apparent 
visual magnitudes of the objects in the sense that larger symbols 
denote brighter objects.

\begin{table}[t]
\begin{center}
\caption{Summary of results; the ages were taken from the literature. In
parenthesis are the errors in the final digits of the corresponding quantity.}
\label{all_res}
\begin{tabular}{lcc}
\hline
\hline
Name & NGC~6192 & NGC~6451 \\
     & C1636$-$432 & C1747$-$302 \\
\hline
$l/b$ & 341/+2 & 360/$-$2 \\
Tr-type & I 2 r & I 2 r \\
$E(b-y)$ & 0.40(5) & 0.40(5) \\
$V_{0}-M_{V}$ & 11.40(10) & 12.15(10) \\
$d$\,[pc] & 1900(100) & 2690(100) \\
\,[Fe/H] & $-$0.10(9) & $-$0.34(6) \\
log\,$t$ & 7.95 & 8.30\\
n(obj) & 374 & 912 \\
n(frames) & 18 & 12 \\
\hline
\end{tabular}
\end{center}
\end{table}

The complete tables with all photometric data for the two cluster stars are available
in electronic form via anonymous ftp 130.79.128.5 or
http://cdsweb.u-strasbg.fr/Abstract.html and  
can be requested from the first author. These tables include the cross
identification of objects from the literature, the observed $(b-y)$,
$m_1$ and $c_1$ values with their corresponding errors, $V$ magnitudes,
as well as the $(B-V)$ and $(U-B)$ values from the literature.

\section{Calibration into the standard photometric system}

\begin{figure*}
\begin{center}
\centerline{\vbox{\psfig{figure=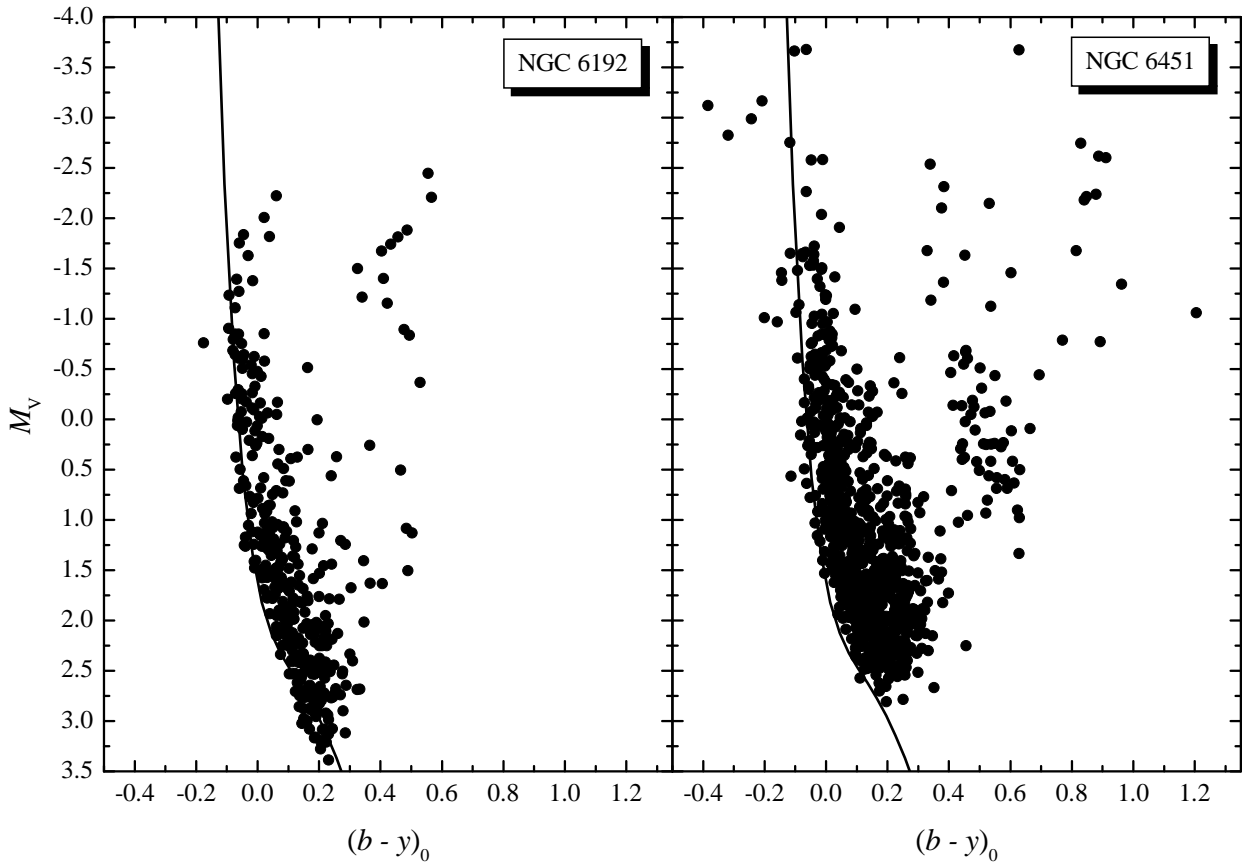,width=170mm}}}
\caption[]{$M_{V}$ versus $(b-y)_0$ diagrams
for our program clusters; the standard relation is taken from
Philip \& Egret (1980).}
\label{all_plot}
\end{center}
\end{figure*}

A very important point is the calibration of the observed photometric
values to standard ones. For the determination of the extinction of the
individual filters, three spectrophotometric standard stars LTT~377,
LTT~1020 and LTT~9239 were observed. Spectrophotometric as well as broadband Johnson
$UBVRI$ photometry for these objects can be found in Hamuy et al. (1992, 1994) and 
Landolt (1992). Unfortunately, no Str\"omgren $uvby$ measurements for these objects are
available. However, we have used the spectrophotometric measurements and the 
standard Str\"omgren $uvby$ transmission curves (Crawford \& Barnes 1970) 
to calculate ``synthetic'' photometric indices for these three standard stars.
The small range of the color indices (e.g. $<$0.1\,mag for $b-y$) makes any reliable 
calibration for the whole relevant color range (more than
1.2\,mag; see Fig. \ref{all_plot}) very problematic. 
However, we used the calculated photometric indices to
derive absolute fluxes using the calibration of Gray (1998). He lists absolute flux
calibrations for the Str\"omgren $uvby$ bands on the basis of data from Vega. The data
for our three standard stars are intrinsic consistent and a comparison with the
the results listed for the Johnson $UBVRI$ by Hamuy et al. (1992, 1994) and 
Landolt (1992) shows an excellent agreement.

For the determination of the extinction coefficients we 
closely follow the approach as described in Sung \& Bessell (2000).
For this purpose we used the measurements
of the three standard stars at airmasses from 1.0 to 2.0 at the two different nights.
The coefficients were found to be 0.48, 0.31, 0.18 and
0.11 for $u$, $v$, $b$ and $y$, respectively.

\begin{figure}
\begin{center}
\centerline{\vbox{\psfig{figure=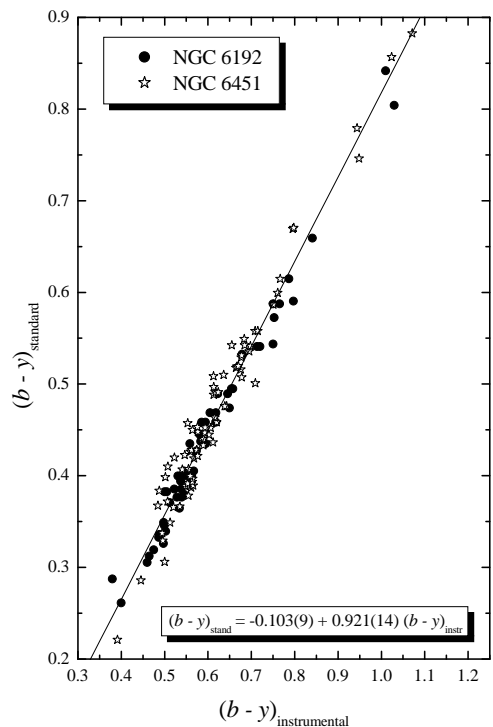,width=70mm}}}
\caption[]{$(b-y)_{standard}$ versus $(b-y)_{instrumental}$ diagram
for our program clusters.}
\label{calib}
\end{center}
\end{figure}

For the calibration of the instrumental $b-y$, $m_1$ and $c_1$ values we have
used the following procedure. The observed CCD photometric $B-V$ values for
NGC~6192 (King 1987, Kjeldsen \& Frandsen 1991) and NGC~6451 (Kjeldsen \& Frandsen 1991)
were transformed into $b-y$ magnitudes using the correlations listed in Caldwell
et al. (1993). These values served as a standard system for the calibration of our
instrumental values. Figure\,\ref{calib} shows the $(b-y)_{stand}$ versus 
$(b-y)_{inst}$ diagram with the available data for both open clusters.
The line fits the data of the two clusters very well. Taking the calibrated
$b-y$ magnitudes, $[m_1]$ and $[c_1]$ indices were calculated using
the absolute fluxes derived for the three standard stars as described above. 
Since these indices
are not sensitive to reddening (Crawford \& Mander 1966) the offsets of the
instrumental $m_1$ and $c_1$ can be estimated. The $y$ magnitudes were directly
converted into standard Johnson $V$ ones using the data of 
King (1987) and Kjeldsen \& Frandsen (1991).

We find the following transformation for the instrumental magnitudes:
\begin{eqnarray}
V &=& -3.02(4) +0.997(2)\cdot y_i \\
b-y &=& -0.103(9) + 0.921(14)\cdot (b-y)_i \\
m_1 &=& -0.723(9)\cdot (m_1)_i \\
c_1 &=& +0.146(8)\cdot (c_1)_i 
\end{eqnarray}
These equations hold for the data of both open clusters.

\section{Program cluster}

In the following two section we will summarize the
already published information about NGC~6192 and NGC~6451.

\subsection{NGC~6192} 

Kilambi \& Fitzgerald (1983; photographic photometry) and King (1987; CCD photometry) 
reported a rather old age (1300\,Myr) with a reddening of $E(B-V)$\,=\,0.26\,mag
and a distance of about 1000\,pc from the Sun. Kjeldsen \& Frandsen (1991; CCD
photometry) on the other hand list
a reddening of $E(B-V)$\,=\,0.68\,mag, an age of 89\,Myr and a distance of
about 1700\,pc from the Sun. The sources of CCD photometry agree very well
and have been used for our analysis.

Paunzen \& Maitzen (2002) analysed all published photometric data and found that
the age, reddening and distance modulus from Kjeldsen \& Frandsen (1991) 
give the best fit. They also
reported the detection of one photometric CP candidate (No.\,54 according to King 1987; 
No.\,86, Kjeldsen \& Frandsen 1991) with $\Delta a$\,=\,+0.030\,mag. From
their photometry a spectral type of B8 Si and an absolute magnitude of 
$M_{\rm V}$\,=\,+0.0(2)\,mag was concluded.

\subsection{NGC~6451}

Paunzen \& Maitzen (2002) investigated the discrepancies of the data
published by Piatti et al. (1998) and Kjeldsen \& Frandsen (1991) for
this open cluster. Piatti et al. (1998) lists $E(B-V)$\,=\,0.08\,mag 
whereas Kjeldsen \& Frandsen (1991)
give $E(B-V)$\,=\,0.70\,mag as best value. This results in a difference of 560\,pc
versus 2100\,pc and 5\,Gyr versus 0.2\,Gyr, respectively. Since
Piatti et al. (1998) calibrated their isochrones using the data for
NGC~6451, we decided to observe this open cluster via Str\"omgren $uvby$ photometry.

Already Paunzen \& Maitzen (2002) concluded that the $V$ magnitudes
given by Piatti et al. (1998) are on the average one magnitude too bright and thus
the reddening, age and distance listed within this reference are probably incorrect.
They refitted the data of Piatti et al. (1998) and found $E(B-V)$\,=\,0.60\,mag,
$(V-M_{\rm V})$\,=\,13.90 and an age of about 0.4\,Gyr which is consistent with
the results of Kjeldsen \& Frandsen (1991).

Furthermore, Paunzen \& Maitzen (2002) found one photometric CP candidate
within the boundaries of NGC~6451.

\section{Results} \label{sres}

First of all we have estimated the reddening of NGC~6192 and NGC~6451.
Taking the standard relation of Philip \& Egret (1980), the observed
magnitudes were shifted accordingly. Figure\,\ref{all_plot} shows the
$M_{V}$ versus $(b-y)_0$ diagrams. The reddening of both program clusters
was found to be $E(b-y)$\,=\,0.40(5)\,mag. This is well in line with the
estimates of Kjeldsen \& Frandsen (1991) and clearly contradicts the values
listed by King (1987; NGC~6192) and Piatti et al. (1998; NGC~6451). For the
distance modulus we find $(V_{0}-M_{\rm V})$\,=\,11.40(10) and 12.15(10) 
resulting in a distance of 1900(100) as well as 2690(100)\,pc
for NGC~6192 and NGC~6451, respectively. Our results agree within 1$\sigma$
(NGC~6192) and 3$\sigma$ (NGC~6451) with those of Kjeldsen \& Frandsen (1991).
Otherwise, several red giants are nicely visible for NGC~6192 whereas
several bright ($M_{\rm V}>-2.5$\,mag) non-members for NGC~6451 were detected.
We have not estimated the ages of both open clusters since no isochrones for
the Str\"omgren $uvby$ system are available. The only way would be the
transformation of the $(b-y)_0$ magnitudes into effective temperature which
is strongly depending on (unavailable) $\beta$ measurements. However,
the ages listed in the literature are compatible with those derived from a
comparison with the isochrones given in Schulz et al. (2002).
Table\,\ref{all_res} lists all results deduced from our analysis.

\begin{figure}
\begin{center}
\centerline{\vbox{\psfig{figure=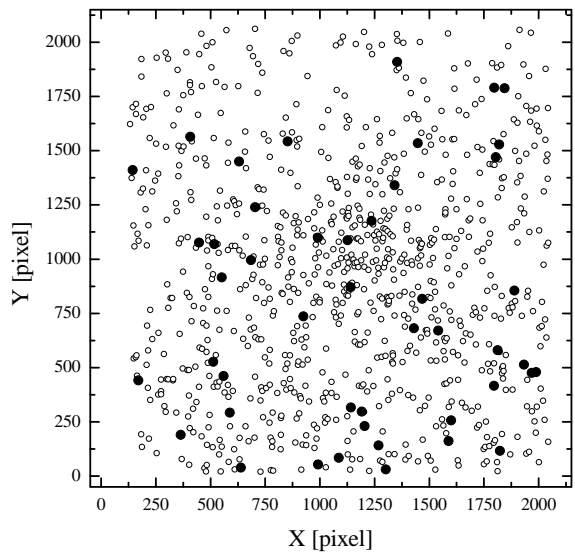,width=80mm}}}
\caption[]{The distribution of apparent members (open circles) of NGC~6451
and the ``shifted population'' as described in Sect.\,\ref{sres}.}
\label{distribution}
\end{center}
\end{figure}

The $M_{V}$ versus $(b-y)_0$ diagram for NGC~6451 shows a clustering of objects
parallel to the main sequence at $0.4 \leq (b-y)_0 \leq 0.7$ and
$+1.0 \leq M_{\rm V} \leq -0.5$. We have investigated the location of these
objects in respect to the field of NGC~6451 (Fig.\,\ref{distribution}). The
placement of these stars
is randomly distributed over the whole observed area ruling out a possible
inconsistency in the reduction process as well as a second, more distant,
open cluster within the vicinity of NGC~6451. Since almost none of these
objects are within the innermost boundaries of NGC~6451, we believe that
the found clustering is a manifestation of the more distant field population.

As next step, the overall metallicities for both open clusters was estimated.
For this purpose the calibration of Nissen (1981) for stars cooler than
F0 was taken. Since no $\beta$ measurements are available we used the standard
relations of Crawford (1975) and Olsen (1984) to estimate them for the
relevant temperature domain. Such a procedure has to be taken with caution.
Applying this method results in [Fe/H]\,=\,$-$0.10(9) and $-$0.34(8) for
NGC~6192 and NGC~6451, respectively. The errors in brackets are the errors
of the means. This seems to indicate that the metallicity for NGC~6451
is slightly subsolar. Another method to calibrate metallicities for red giants
is listed by Hilker (2000). This calibration is valid for objects with
0.5\,$<$\,$(b-y)_0$\,$<$1.0\,mag. This fact together with the apparent uncertainty
of the membership of objects for NGC~6451 seriously limits the significance
of the derived values. Only two and five objects remain for NGC~6192
and NGC~6451, respectively which results in errors as large as $\pm$0.5\,dex.

The investigation of the photometric candidate CP star in NGC~6192 (Paunzen
\& Maitzen 2002) further strengthens its chemically peculiar nature. 
We find $(b-y)_0$\,=\,$-$0.026, [$m_1$]\,=\,0.202, [$c_1$]\,=\,0.738 and
$M_{V}$\,=\,0.21\,mag which are typical for a B8\,Si star (Cameron 1966).

The values for the apparent peculiar object in NGC~6451 are typical for
an unreddened G-type star (Crawford \& Mander 1966) and rule out a 
chemical peculiarity: $(b-y)$\,=\,0.651, [$m_1$]\,=\,0.467, [$c_1$]\,=\,0.233 
and $M_{V}$\,=\,3.85\,mag. This case shows the importance of further
photometric or spectroscopic investigations of candidate
CP stars.

\section{Conclusions} 

We have presented Str\"omgren $uvby$ photometry for the young
open clusters NGC~6192 (374 objects) and NGC~6451 (912 objects). 
For both open clusters
widely different reddening values and thus distances and ages
can be found in the literature. From our photometry we were able
solve these discrepancies and derive new accurate values. The overall
metallicities was found to be about solar for NGC~6192 as well as
slightly subsolar for NGC~6451 using the appropriate calibrations
from the literature. 

The chemically peculiar nature for one object of NGC~6192 was
proven whereas one photometric CP candidate within NGC~6451 has to be
rejected.

\begin{acknowledgements}
EP acknowledges partial support by the Fonds zur F\"orderung der
wissenschaftlichen Forschung, project P14984.
Use was made of the SIMBAD database, operated at CDS, Strasbourg, France and
the WEBDA database, operated at the Institute of Astronomy of the University
of Lausanne.
\end{acknowledgements}

\end{document}